\documentclass[amsmath,amsfonts,prx,twocolumn,longbibliography]{revtex4-1}
\usepackage{graphicx}
\usepackage{epsfig}
\usepackage{bbm}
\usepackage{bm}
\usepackage{dcolumn}
\usepackage{amsmath}
\usepackage{amssymb}
\usepackage{color}

\newcommand{\beg}{\begin{equation}}
\newcommand{\en}{\end{equation}}
\newcommand{\bp}{\mathbf p}
\newcommand{\bq}{\mathbf q}
\newcommand{\bk}{\mathbf k}

\newcommand \bel  {\begin{align}}
\newcommand \enl  {\end{align}}

\newcommand{\veps}{\varepsilon}
\newcommand{\eps}{\epsilon}

\newcommand{\up}{\uparrow}
\newcommand{\dn}{\downarrow}
\newcommand{\dg}{^\dagger}

\begin{document}

\title{Migdal-Eliashberg superconductivity in a Kondo lattice}

\author{Samuel Awelewa}
\affiliation{Department of Physics, Kent State University, Kent, OH 44242, USA}

\author{Maxim Dzero}
\affiliation{Department of Physics, Kent State University, Kent, OH 44242, USA}

\begin{abstract}
We apply the Migdal-Eliashberg theory of superconductivity to heavy-fermion and mixed valence materials. Specifically, we extend the Anderson lattice model to a case when there exists a strong coupling between itinerant electrons and lattice vibrations. Using the saddle-point approximation, we derive a set of coupled nonlinear equations which describe competition between the crossover to a heavy-fermion or mixed-valence regimes and conventional superconductivity.
We find that superconductivity at strong coupling emerges on par with the development of the many-body coherence in a Kondo lattice. Superconductivity is gradually suppressed with the onset of the Kondo screening and for strong electron-phonon coupling the Kondo screening exhibits a characteristic re-entrant behavior. Even though for both weak and strong coupling limits the suppression of superconductivity is weaker in the mixed-valence regime compared to the local moment one, superconducting critical temperature still remains nonzero. In the weak coupling limit the onset of the many body coherence develops gradually, in the strong coupling limit it emerges abruptly in the mixed valence regime while in the local moment regime the $f$-electrons remain effectively decoupled from the conduction electrons. Possibility of experimental realization of these effects in Ce-based compounds is also discussed. 
\end{abstract}


\date{\today}

\maketitle

\section{Introduction}
Since the discovery of the first heavy-fermion system CeAl$_3$ \cite{CeAl3} along with the subsequent discoveries of unconventional superconductivity in heavy-fermion compounds UBe$_{13}$ \cite{UBe13} and CeCu$_2$Si$_2$ \cite{Ce122}, which serve as the very first examples of this remarkable phenomenon in solid state systems, heavy-fermion materials continue to provide an indispensable platform for developing novel physical concepts \cite{Varma1976,hilbert2,rise,Mydosh1985,Hewson,Coleman2007,Qi-RMP20,BarzykinTwoFluid}. In heavy-fermion materials strong hybridization between itinerant and quasilocalized $f$-electrons leads to an emergence of competing interactions as well as drastic changes in the single-particle properties, including an increase of the effective mass of the carriers exceeding by several orders of magnitude bare electron mass, $m_{\textrm{eff}}\sim 100\div1000 m_e$. Renewed interest to these systems has been motivated by the discovery of the topologically protected metallic surface states \cite{Dzero2016} and quantum oscillations in samarium hexaboride \cite{Suchitra2017}, discovery of the topological Weyl-Kondo semimetals \cite{WeylKondo2018} and, most recently, experimental and theoretical studies of unconventional superconductivity in UTe$_2$ \cite{Aoki2022,UTe2Dan,UTe2Ref2,UTe2Ref3}.

Emergence of unconventional superconductivity in heavy fermion systems is usually attributed to either existence of the enhanced magnetic fluctuations due to partially screened magnetic moments \cite{Coleman2007,Qi-RMP20,Cedo2001,Sarrao2007} or due to strong valence fluctuations on $f$-sites \cite{Miyake_2007,Miyake2000,KVC1,KVCErr}. However, there is a number of Ce- or Yb-based heavy-fermion systems such as CeRu$_3$Si$_2$ and its alloys Ce$_{1-x}$La$_x$Ru$_3$Si$_2$, CeRu$_2$, CeNi$_2$Ge$_2$ in which superconductivity seems to be 'inherited' from their respective homolog counterparts LaRu$_2$ and LaNi$_2$Ge$_2$ correspondingly \cite{Hull1981,Wilhelm2002,Maezawa1999,Gegenwart1999,Kishimoto2003,Kishimoto2003b}. Since homologs do not contain atoms with partially filled $f$-orbitals, superconductivity in these materials is most likely conventional and, therefore, must be mediated by the electron-phonon interactions. 

To the best of our knowledge, Barzykin and Gor'kov were the first ones to address the question of whether the phonon-mediated superconductivity can be completely excluded as a possible microscopic mechanism of superconductivity in heavy-fermion materials \cite{LPG2005}. They have employed weak-coupling (BCS) theory \cite{bcstheory} and studied the conditions for the emergence of the superconducting state under assumption that Kondo lattice coherence temperature is larger than critical temperature of the superconducting transition. It was found that in a narrow range of the model parameters describing the $f$-electron subsystem, superconducting critical temperature remains finite. In other words, already within the framework of the BCS theory the short answer to the question above is negative.
 
In this paper we will generalize Barzykin-Gor'kov theory to a case when electron-phonon coupling is strong. In particular, we consider the emergence of superconductivity at strong coupling along with the emergence of the heavy-fermions on equal footing. As it was early noted by Eliashberg \cite{EliashbergLetter}, the hybridization between itinerant and predominantly localized $f$-electrons leads to renormalization of the spectrum of the conduction electrons and is analogous to a Migdal renormalization \cite{Migdal} due to electron-phonon interaction. It is therefore of special interest to us to investigate an interplay between these two physical phenomena without making any specific assumptions on the relative values of the hybridization and strength of the electron-phonon coupling. 

The present work has also been motivated in part by the recent experimental results on Ce-based cage compounds CeNi$_2$Cd$_{20}$ and CePd$_2$Cd$_{20}$ which do not exhibit long-range order down to a millikelvin temperature range \cite{White2015,DzeroRKKY2023}. Given the large separation between Ce ions, both the super-exchange interaction and Ruderman–Kittel–Kasuya–Yosida (RKKY) interaction are \cite{Ruderman1954,Kasuya1956,Yosida1957} are substantially suppressed which is supported by the analysis of the temperature dependence of magnetic susceptibility. Since in these systems the Kondo lattice coherence temperature is also fairly low, upon further cooling these systems may also develop superconductivity mediated by the electron-phonon interaction. 

This paper is organized as follows. In the next Section we formulate a problem and provide the qualitative discussion based on the recently developed pseudospin formulation of Migdal-Eliashberg theory. In Section III we formulate the microscopic model and derive the main equations within the saddle-point approximation for the effective action. Section IV is devoted to the discussion of the results which follow from the self-consistent solution of the saddle-point equations. In Section V we provide the main conclusions that one can draw from the present work. Necessary technical details, which are needed to follow the technical part of this paper, are summarized in Appendix A.
 
\section{Qualitative discussion}
A problem of an interplay between $s$-wave superconductivity and Kondo screening \cite{Kondo64} dates back to 1970s \cite{MHZ1970,MHZ1971}. In a dense Kondo lattice an interplay between BCS superconductivity and Kondo screening of magnetic impurities leads to a phenomenon of re-entrant superconductivity: at when superconducting critical temperature $T_c\gg T_K$ ($T_K$ is a single impurity Kondo temperature) 
superconductivity is suppressed as the number of impurities is increasing. When $T_c$ becomes comparable to $T_K$ suppression becomes so strong that superconductivity is destroyed until it recovers at much lower $T_c\ll T_K$ since at that temperatures the paramagnetic moments are fully screened.  It is worth mentioning that similar phenomenon of re-entrant superconductivity was also discussed in the context of the charge Kondo alloys Pb$_{1-x}$Tl$_x$Te \cite{Dzero2005,Fisher2005}.

Let us now turn our attention to a case of superconductivity mediated by strong electron-phonon interaction commonly known as Migdal-Eliashberg theory \cite{Eliashberg}. It has been recently shown that Migdal-Eliashberg theory can be elegantly formulated in terms of the normalized pseudospin variables ${\vec S}_n=(S_n^x,S_n^y,S_n^z)$ in the Matsubara frequency representation  with the components which satisfy the normalization condition ${\vec S}_n^2=1$ \cite{EmilEli1}. The subscript refers to the fermionic Matsubara frequency $\omega_n=\pi T(2n+1)$ $(n=0,\pm1,\pm2, ...)$. If one considers the effects of the Kondo lattice within the large-$N$ approximation than the results of \cite{EmilEli1} can be easily generalized for the case of a Kondo lattice as well. The corresponding expression for the Hamiltonian is (see Appendix A):
\beg\label{KLSpins}
\begin{split}
{H}_{\textrm{eff}}&=-2\sum\limits_{n}\omega_n\left(1+\frac{v^2}{\omega_n^2+\veps_f^2}\right)S_n^z\\&-\overline{g}T\sum\limits_{nm}\lambda_{nm}(S_n^xS_m^x+S_n^zS_m^z).
\end{split}
\en
Here $\lambda_{nm}=[(\omega_n-\omega_m)^2/\Omega^2+1]^{-1}$, $\Omega$ is the frequency of the optical phonon, $\overline{g}$ is dimensionless parameter determined by the strength of the electron-phonon coupling, parameter $v$ is the hybridization with the Kondo lattice and $\veps_f$ is the single particle energy of a localized $f$-electron renormalized by the hybridization between the conduction and $f$-electrons ($\veps_f$ is usually associated with the Kondo lattice coherence temperature $T_K$ \cite{Miyake2000,Panday2023}). Clearly, as it follows from the first term (\ref{KLSpins}) in the spin representation Kondo lattice effects amount to the renormalization of the Matsubara frequencies. 

We can now consider the case when all $f$-electrons are occupying states with energy $\eps_{f0}$ (i.e. $f$-levels form a flat band). These levels are all singly occupied, so we assume that local moments are formed and we will also assume that interactions between the local moments are extremely weak \cite{White2015,DzeroRKKY2023}.  As temperature is lowered, hybridization between the conduction and $f$-electrons, which ultimately leads to an onset of the many-body coherence, competes with the superconducting pairing. When $\overline{g}$ is small enough, the system is in the weak coupling regime and naturally when $v$ is increasing the superconductivity will be gradually suppressed \cite{LPG2005}. With an increase in $\overline{g}$, however, the situation analogous the one for the Kondo impurity, which we discussed above, may arise. Specifically, for a given value of $\overline{g}$ we may find two distinct configurations ${\vec S}_n$ each realized for a given pair of $v$ and $\veps_f$: one would correspond to a state with $v/\Omega\ll 1$, while another to a case when $v/\Omega\sim 1$. 

It is worth mentioning here that in the context of the underlying interactions between the localized and itinerant electrons, there several regimes which must be distinguished. In the Kondo lattice regime $\eps_{f0}$ lies well below the bottom of the conduction band and, as a consequence the transitions between the $f$ states and conduction band states are purely virtual. In this regime $|\eps_{f0}|\gg\nu_FV^2$, where $\nu_F$ is the single particle density of states at the Fermi level and $V$ is the hybridization amplitude, so that parameter $v\ll 1$. In the mixed valence regime $|\eps_{f0}|\sim\nu_FV^2$ and $v\sim 1$. We therefore expect that two distinct spin configurations will most likely exist in the mixed valence regime, since in this regime the $f$-electron occupation number may significantly decrease below its value in the Kondo regime $n_f\sim1$. Furthermore, since there are no 'hidden' symmetries in the problem, it is clear that free energies corresponding to two solutions will differ. This implies that the onset of the screening in the Kondo lattice may develop abruptly as the first-order transition. In what follows below by using the saddle-point approximation for the Anderson lattice model with electron-phonon interactions, we will show that indeed, in the mixed valence regime there exists a possibility for a $s$-wave superconductivity at strong coupling to emerge from the many-body coherent state which is, in turn, formed by a strong hybridization between the conduction and localized $f$-electons.

\section{Model}
We consider a system which includes itinerant electrons interacting with predominantly localized electrons and also with the lattice vibrations.  The model Hamiltonian can be written as a sum of two terms 
\beg\label{Eq1}
{\cal H}={\cal H}_{\textrm{ALM}}+{\cal H}_{\textrm{e-ph}}.
\en
The first term is the Anderson lattice model Hamiltonian which describes the interactions between the itinerant ($c$) and quasilocalized ($f$) electronic degrees of freedom:
\beg\label{Eq2}
\begin{split}
{\cal H}_{\textrm{ALM}}&=\sum\limits_{\bk\sigma}\eps_\bk c_{\bk\sigma}\dg c_{\bk\sigma}+\eps_{f0}\sum\limits_{\bk\sigma}f_{\bk\sigma}\dg f_{\bk\sigma}\\&+V\sum\limits_{\bk\sigma}\left(f_{\bk\sigma}\dg c_{\bk\sigma}+c_{\bk\sigma}\dg f_{\bk\sigma}\right)+U_{\textrm{ff}}\sum\limits_in_{i\up}^f n_{i\dn}^f.
\end{split}
\en
Here $c_{\bk\sigma}\dg(c_{\bk\sigma})$ and $f_{\bk\sigma}\dg (f_{\bk\sigma})$ are the fermionic creation (annihilation) operators for the conduction and $f$-electrons with momentum $\bk$ and spin projection $\sigma$ correspondingly, $\eps_\bk=k^2/2m-D$ is the single-particle energy for the conduction electrons taken relative to the half of the conduction band width $D$, $m$ is an electron's mass, $\eps_{f0}$ single particle energy for the localized electrons, $V$ is the hybridization amplitude which without loss of generality is assumed to be local and $n_{i\sigma}^f=f_{i\sigma}\dg f_{i\sigma}$. Lastly, $U_{\textrm{ff}}$ is the local Coulomb interaction between the $f$-electrons and we assume that $U_{\textrm{ff}}\gg \textrm{max}\{D,V\}$. 
The second term in (\ref{Eq1}) accounts for the lattice excitations and their interactions with the conduction electrons:
\beg\label{Eq1eph}
{\cal H}_{\textrm{e-ph}}=
\sum\limits_\bp \omega_\bp{b}_\bp\dg{b}_\bp+g\sum\limits_{\bp\bq,\sigma}
{c}_{\bp+\bq,\sigma}\dg {c}_{\bp\sigma}\left({b}_{-\bq}\dg+{b}_\bq\right).
\en
Here $\omega_\bp$ is the phonon dispersion, $b_\bq\dg$, $b_\bq$ are bosonic creation and annihilation operators and $g$ is the electron-phonon coupling constant.

Since the repulsion between the $f$-electrons serves as the largest energy scale in the problem, we will consider the limit $U_{\textrm{ff}}\to\infty$. In this case one can omit the last term in (\ref{Eq2}) and introduce the projection (slave-boson) operators by replacing 
$f_{j\sigma}\to f_{j\sigma}p_j\dg$ \cite{read83a,read83b,me1983,PiersLong,coleman_optical,PiersLong,auerbach}. In addition to having slave-boson operators, one needs to introduce the constraint
\beg\label{Qj}
Q_j=\sum\limits_{\sigma=1}^2f_{j\sigma}\dg f_{j\sigma}+p_j\dg p_j=1.
\en
This constraint is needed in the large-$N$ formulation of the theory to keep the same size of the Hilbert space as in the original one with spin-$1/2$ particles as well as to ensure that $f$-sites are not doubly occupied. In addition, in what follows we will consider a case of an interaction with the single Einstein (optical) harmonic oscillator with frequency $\Omega$.

\begin{figure}
\includegraphics[width=0.835\linewidth]{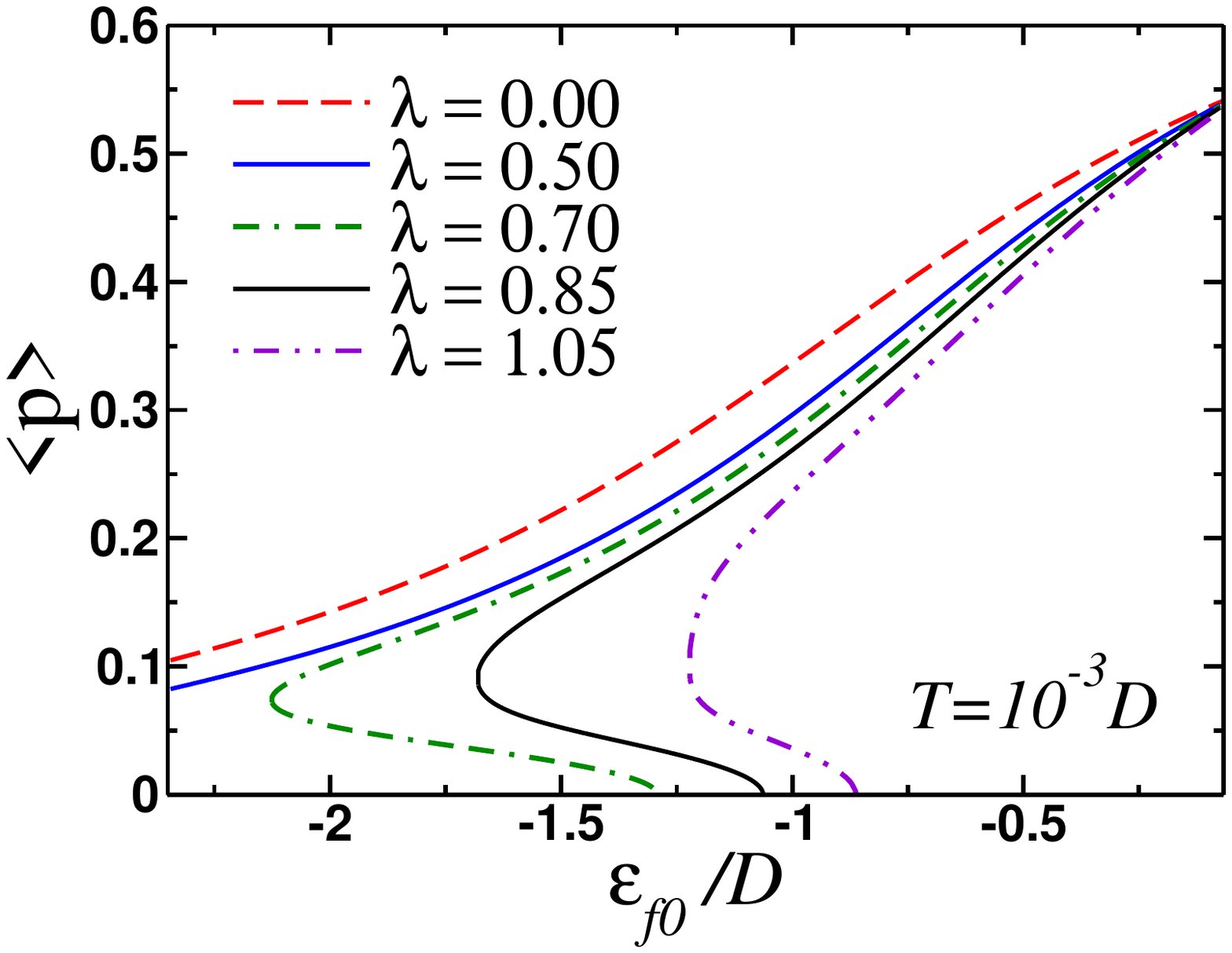}
\caption{Results of the self-consistent solution of the saddle-point equations (\ref{EliKLFin},\ref{SBEqs})
for the slave-boson amplitude $\langle p\rangle$ as a function of the bare $f$-electron energy $\eps_{f0}$. When the electron-phonon coupling is strong enough we find that there are two regimes in which superconductivity co-exists with the onset of the Kondo lattice coherence: in the first regime the conduction electrons are weakly coupled to $f$-electrons ($\langle p\rangle\ll 1$), while in the second regime the coupling between conduction and $f$-electrons is much stronger $(\langle p\rangle\sim 1/2)$. The existence of these two regimes is very similar to the phenomenon of the re-entrant superconductivity in dense Kondo alloys. As it follows from our calculation, two solutions for $\langle p\rangle$ and $\Delta_n$ for the same value $\eps_{f0}$ are possible provided that $\lambda \sim 1$ and also when the system is neither in the local moment regime, $|\eps_{f0}|\gg \pi\nu_F|V|^2$, nor in the mixed valence regime, $|\eps_{f0}|\ll \pi\nu_F|V|^2$. }
\label{Fig-ReEnter}
\end{figure}

\subsection{Saddle-point approximation}
With these provisions we may follow the avenue of Refs. \cite{Miyake2000,Panday2023,EmilEli1,EmilEli2} to derive a set of saddle-point equations in the Matsubara representation. These equations are obtained by finding an extremum of the effective action (see Appendix \ref{EffectiveAction} for details).  The Eliashberg equations for the Matsubara components of the pairing field $\Phi(\tau)$ and self-energy $\Sigma(\tau)$ are
\beg\label{EliHF}
\begin{split}
\Phi_n&=\frac{T}{\pi \nu_F}\sum\limits_{\omega_m}\sum\limits_{\bk}\frac{D(\omega_n-\omega_m)\Phi_m}{
\omega_m^2Z_m^2+(\xi_\bk-\veps_f\sigma_m)^2+|\Phi_m|^2}, \\
\Sigma_n&=\frac{T}{\pi \nu_F}\sum\limits_{\omega_m}\sum\limits_{\bk}\frac{D(\omega_n-\omega_m)\omega_mZ_m}{
\omega_m^2Z_m^2+(\xi_\bk-\veps_f\sigma_m)^2+|\Phi_m|^2}.
\end{split}
\en
Here $\omega_mZ_m=\omega_m(1+\sigma_m)+\Sigma_m$, $\nu_F$ is the single-particle density of states at the Fermi level, $\omega_n=\pi T(2n+1)$ are fermionic Matsubara frequencies, $\xi_\bk=\eps_\bk-\mu$, $\mu$ is the chemical potential, 
$\sigma_m=a_{\textrm{sb}}^2/(\omega_m^2+\veps_f^2)$, $a_{\textrm{sb}}=V\langle p\rangle$, $\veps_f$ is the renormalized position of the $f$-electron energy level and $D(\omega_n)=
{\lambda\Omega^2}/(\omega_n^2+\Omega^2)$
is the propagator of an optical phonon with $\lambda$ being dimensionless electron-phonon coupling constant \cite{Carbotte1,Carbotte2,Combescot}. In equations (\ref{EliHF}) we have neglected the effects associated with absence of the particle-hole symmetry. In fact, we have verified that this approximation does not affect our subsequent results in any substantial way.

In equations (\ref{EliHF}) we can perform an integration over $\xi_\bk$ by extending the limits to infinity, which yields
\beg\label{EliKLFin}
\begin{split}
\Phi_n&=T\sum\limits_{\omega_m}\frac{D_{nm}\Phi_m}{\sqrt{[\omega_m(1+\sigma_m)+\Sigma_m]^2+|\Phi_m|^2}}, \\
\Sigma_n&=T\sum\limits_{\omega_m}\frac{D_{nm}[\omega_m(1+\sigma_m)+\Sigma_m]}{\sqrt{[\omega_m(1+\sigma_m)+\Sigma_m]^2+|\Phi_m|^2}},
\end{split}
\en
where $D_{nm}=D(\omega_n-\omega_m)$.
We immediately note that the Kondo lattice lattice effects enter into these equations through the function $\sigma_n$ only, so that the equation for $\Phi_n$ has exactly the same form as in the Migdal-Eliashberg theory. Consequently, in order to analyze these equations we use the following parametrization for the pairing fields: $\Phi_m=\Delta_mZ_m$.

\begin{figure}
\includegraphics[width=0.85\linewidth]{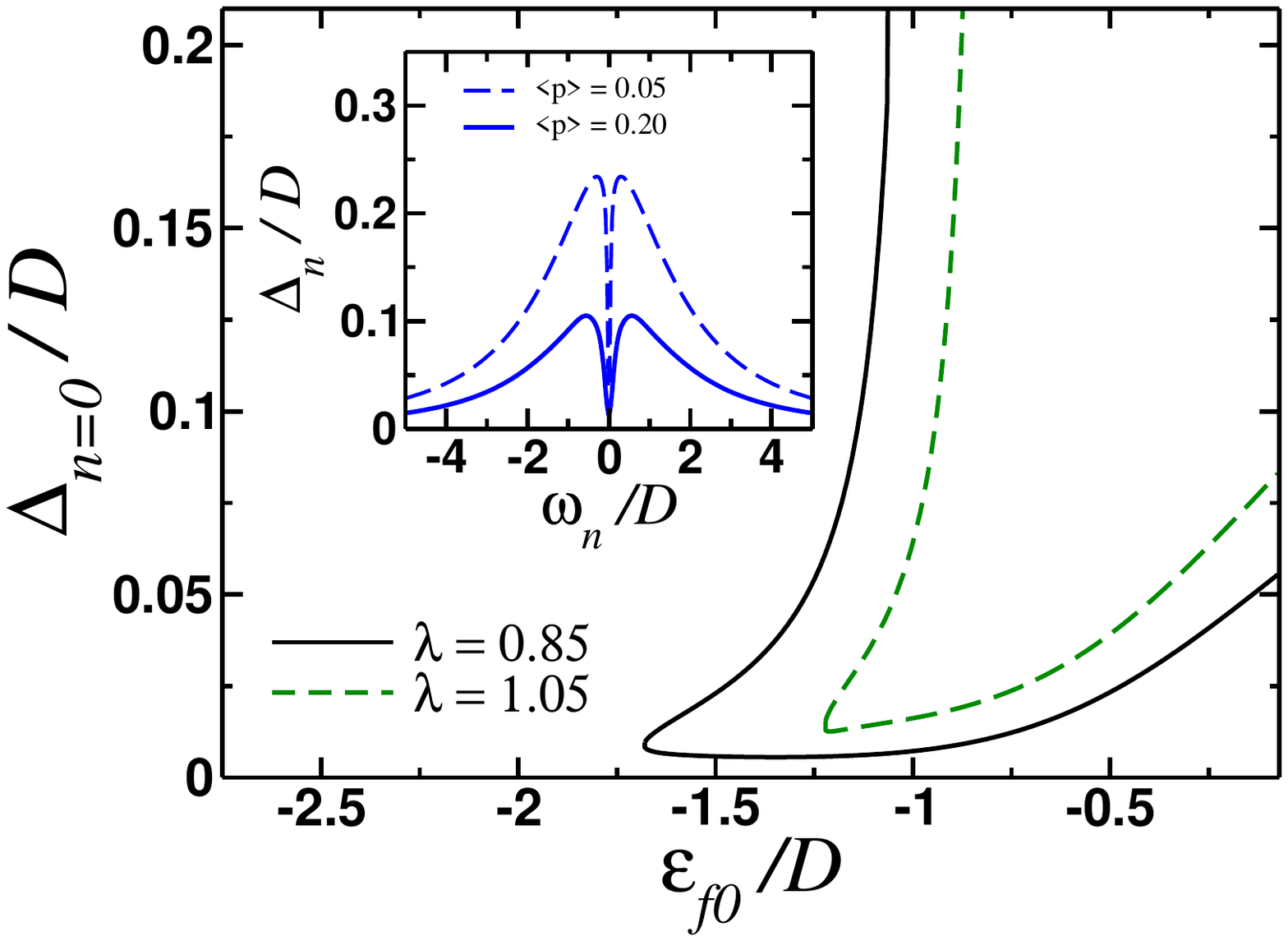}
\caption{Results of the self-consistent solution of the saddle-point equations (\ref{EliKLFin},\ref{SBEqs})
for the pairing function $\Delta_{n=0}$ computed for $\omega_0=\pi T$ (main panel). The inset shows the dependence of $\Delta_n$ on $\omega_n$ for two different values of the slave-boson amplitude corresponding to two solutions with $\eps_{f0}=-1.105D$ and $\lambda=1.05$.}
\label{Fig-DLTn}
\end{figure}

If the values of $a_{\textrm{sb}}$ and $\veps_f$ are known then equations (\ref{EliKLFin}) can be solved by iterations. 
Both parameters $\veps_f$ and $a_{\textrm{sb}}$ are computed self-consistently from the slave-boson mean-field equations (see Appendix A for details):
\beg\label{SBEqs}
\begin{split}
\veps_f-\eps_{f0}&=\sum\limits_\bk\frac{n_F(E_{2\bk})-n_F(E_{1\bk})}{\sqrt{({\eps}_\bk-{\varepsilon}_f)^2+4{a}_{\textrm{sb}}^2}}+\delta{\cal F}_1[{\vec \Sigma},{\vec \Phi}], \\
n_{\textrm{f}}-n_{\textrm{c}}&=\sum\limits_\bk\frac{({\eps}_\bk-{\varepsilon}_f)[n_F(E_{2\bk})-n_F(E_{1\bk})]}{\sqrt{({\eps}_\bk-{\varepsilon}_f)^2+4{a}_{\textrm{sb}}^2}}\\
&+\delta{\cal F}_2[{\vec \Sigma},{\vec \Phi}], \\
n_{\textrm{f}}+n_{\textrm{c}}&=\sum\limits_\bk\left[n_F(E_{1\bk})+n_F(E_{2\bk})\right]+\delta{\cal F}_3[{\vec \Sigma}, {\vec \Phi}],
\end{split}
\en
where we have omitted the normalization pre-factors in front of momentum summations for brevity, ${\vec \Sigma}$ and ${\vec \Phi}$ are vectors with the Matsubara components $\Sigma_n$ and $\Phi_n$ correspondingly, $n_F(\eps)=\{\exp[(\eps-\mu)/T]+1\}^{-1}$ is the Fermi distribution function and we introduced
\beg\label{E12k}
E_{1(2)\bk}=\frac{1}{2}\left({\eps}_\bk+{\varepsilon}_f\pm\sqrt{({\eps}_\bk-{\varepsilon}_f)^2+4a_{\textrm{sb}}^2}\right).
\en 
Note that in (\ref{E12k}) we take the value of $\veps_f$ relative to the chemical potential, i.e. we formally replace $\veps_f\to\veps_f-\mu$. Functions $\delta{\cal F}_a={\cal F}_a[{\vec \Sigma},{\vec \Phi}]-{\cal F}_a[{\vec \Sigma}=0,{\vec \Phi}=0]$ vanish identically when $\Sigma_n=\Phi_n=0$. 
Lastly, the chemical potential needs to be computed self-consistently from the particle number conservation $n_{\textrm{c}}+n_{\textrm{f}}=n_{\textrm{tot}}$ (see Appendix \ref{EffectiveAction}). 

The effective mass $m=(3\pi^2/\sqrt{2})^{2/3}/2{\cal D}$ of the conduction electrons is obtained from the condition that the total number of particles (per spin) equals one:
\beg\label{IntPart}
\int\limits_{-{\cal D}}^{{\cal D}}\nu(\eps_\bk)d\eps_\bk=1.
\en
In this formula $\nu(\eps)=(3/4\sqrt{2}{\cal D})\sqrt{\eps/{\cal D}+1}$ is the single-particle density of states for the non-interacting system and ${\cal D}$ is the half-width of the conduction band. In what follows we will limit out calculations to the case when the particle occupation number of the conduction band (per spin) equals to $0.375$,  so that the chemical potential will remain in the lower of the two hybridized bands and the total particle occupation number per spin will be fixed to $n=0.875$. Therefore, in order to compute the momentum summations we will use the following expression:
\beg\label{Sumk}
\sum\limits_{\bk}(...)=\int\limits_{-{\cal D}}^{\cal D}\nu(\eps_\bk)(...)d\eps_\bk.
\en
Thus, in order to determine the system's ground state, we have to solve five nonlinear equations self-consistently.
\begin{figure}
\includegraphics[width=0.8\linewidth]{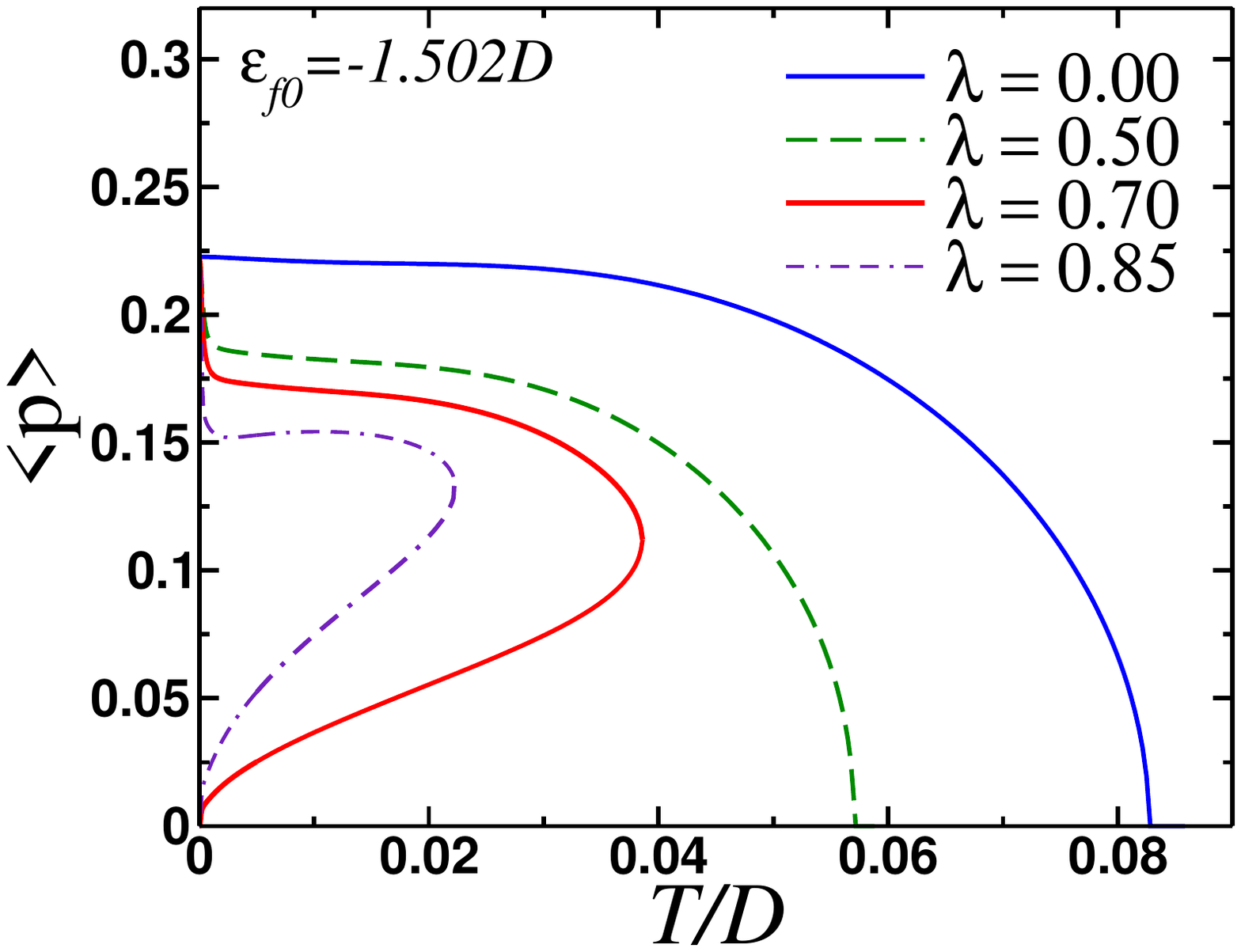}
\caption{Results of the self-consistent solution of the saddle-point equations (\ref{EliKLFin},\ref{SBEqs})
for the slave boson amplitude as a function of temperature. Notably, with an increase in the values of the electron-phonon coupling, the Kondo lattice coherence may develop in the mixed-valence regime only, while the crossover to the local moment regime is strongly suppressed.}
\label{Fig-pvsT}
\end{figure}
\section{Solution of the saddle-point equations}
We solve the system of saddle-point equations (\ref{EliKLFin}) by iterations for a given set of the slave-boson parameters which are used at each step to solve (\ref{SBEqs}). In our numerical calculations we purposefully choose all the remaining energy scales to be of the same order, i.e. ${\cal D}\approx{\Omega}\approx V$. 
We present our results in Fig. \ref{Fig-ReEnter}. We find that when $\lambda \sim 1$ there is a region in parameter space when there two sets of solutions. The first set corresponds to a smaller slave boson amplitude while for the second set the slave boson amplitude has much larger values. Remarkably, when $\lambda \sim 1$, the slave-boson mean-field equations do not have solution in the parameter range corresponding to the local moment regime and it appears as if at very low temperatures two electronic subsystems are completely decoupled from each other. On the other hand, when $|\eps_{f0}|\ll\pi\nu_F|V|^2$, we see that the solution for $\lambda\not=0$ is close to the one for the normal metal, i.e. deep in the mixed-valence regime superconductivity will be completely suppressed by the valence fluctuations. 

In Fig. \ref{Fig-DLTn} we present fully self-consistent solution of the Eliashberg equation for the function $\Delta_n$. We observe that a state with higher maximum value of $\Delta_n$ corresponds to the case when $\langle p\rangle$ is small, which is expected. Note that $\Delta_n$ reaches its maximum value for $n\not=0$ and in the dependence of $\Delta_{n=0}$ on $\eps_{f0}$ we verified that it indeed approaches its value for $V=0$.

In order to get further insight into the nature of the ground state, we have computed the dependence of the slave boson amplitude $\langle p\rangle$ on temperature for $\eps_{f0}\approx -3{\cal D}/2$ which corresponds to the boundary region between the local moment and mixed valence regime. The results are presented in Fig. \ref{Fig-pvsT}. We observe that for small values of the electron-phonon coupling we find a typical temperature dependence of $\langle p\rangle^2\sim 1-T/T_{\textrm{coh}}$, where $T_{\textrm{coh}}$ is the Kondo lattice coherence temperature. However, as the value of the electron-phonon coupling is increased, in the fairly wide range of values of the bare $f$-energy level $\eps_{f0}$ we find that temperature dependence of $\langle p\rangle$ has a characteristic nonmonotonic shape: at low temperatures the slave-boson condensation may happen with both small ($\langle p\rangle\ll 0.1$) and large values ($\langle p\rangle \sim 0.1$) of the amplitude. When $\lambda\sim 1$ the condensation will only be possible when the system will be in the mixed-valence regime ($\langle p\rangle > 0.1$), since there are no solutions for $\langle p\rangle$ when $|\eps_{f0}|\gg \pi\nu_F|V|^2$ at low temperatures and also $n_f$ will significantly deviate from an integer value. 
We may also interpret our results in Fig. 3 as a suddent emergence of the many-body coherence when the value of the slave-boson amplitude changes abruptly from zero to some finite value. 

\begin{figure}
\includegraphics[width=0.8\linewidth]{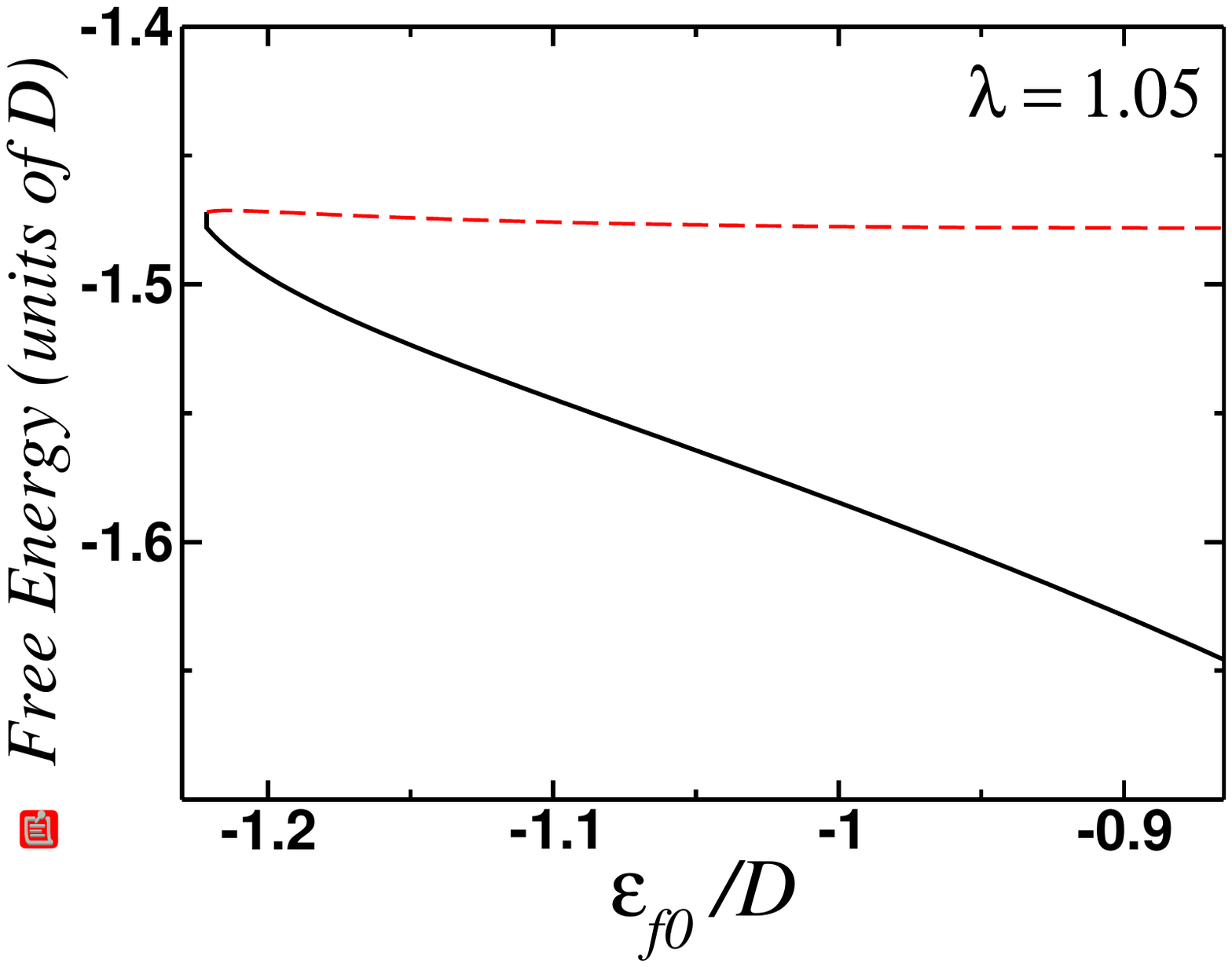}
\caption{Free energy (\ref{FreeEnergy}) dependence on the position of the $f$-electron energy level $\eps_{f0}$ plotted in the region corresponding to two solutions of the self-consistency equations (Fig. 1) for $\lambda=1.05$ and $T=0.005D$. The black solid line corresponds to the upper branch, while the red dashed line corresponds to the lower branch. As expected, the lower (i.e. 're-entrant') branch of that solution has a higher free energy. This result implies that development of the many-body coherence will happen abruptly rather than gradually.}
\label{Fig-FE}
\end{figure}

Existence of the two solutions for the fixed value of $\eps_{f0}$ brings forth the question of whether for these solutions the free energy will have the same value or not. To answer this question we derived the following expression for the free energy, which is given relative to the free energy of the normal state with no hybridization between the conduction and $f$-electrons:
\beg\label{FreeEnergy}
\begin{split}
&\delta F_{\textrm{sc-kl}}=2\left(\langle p\rangle^2-\frac{1}{2}\right)(\veps_f-\eps_{f0})\\&+T\sum\limits_{\bk m}\frac{Z_m(\omega_m\Sigma_m+Z_m\Delta_m^2)}{(\omega_m^2+\Delta_m^2)Z_m^2+(\xi_\bk-\veps_f\sigma_m)^2}\\&-T\sum\limits_{\bk m}\log\left(\frac{\omega_m^2+\veps_f^2}{\omega_m^2+\eps_{f0}^2}\right)\\&\\&-T\sum\limits_{\bk m}\log\left[\frac{(\omega_m^2+\Delta_m^2)Z_m^2+(\xi_\bk-\veps_f\sigma_m)^2}{\omega_m^2+\xi_\bk^2}\right].
\end{split}
\en
The first two terms in this expression describe the free energy due to bosonic degrees of freedom. The third one is the change in the energy of the $f$-electrons due to hybridization with the conduction band, while the last term accounts for the change in the free energy of the conduction electrons. In Fig. \ref{Fig-FE} we show the dependence of the free energy correction, Eq. (\ref{FreeEnergy}), as a function of $\eps_{f0}$ in the region when we find two solutions of the self-consistency equations. 
This means that when the values of the electron-phonon coupling are strong enough and the position of the bare $f$-level is closer to the mixed-valence region, the many-body coherence in the Kondo lattice will indeed develop abruptly, i.e. similar to the iso-structural valence transitions in metallic cerium and YbInCu$_4$ \cite{Lawrence_1981,AKZvezdin2000}. One however needs to keep in mind that the slave-boson condensation in the Kondo lattice is a crossover phenomenon and not a true phase transition.

\subsection{Solution for real frequencies}
In order to determine the dependence of the pairing gap on $\eps_{f0}$ and temperature we need to either solve the Eliashberg equations for real frequencies by using the method of analytic continuation from Matsubara to real frequencies or use P\'{a}de approximation. Given the form of the Eliashberg equations, we will use the analytic continuation since it seems to us as more straightforward. The pairing gap $\Delta$ will then be given by the root of equation
$\omega=\Delta(\omega)$. 
Following the procedure outlined in Refs. \cite{Carbotte1,Combescot}, we immediately observe that the equation for $\Delta(\omega)$ in our case will coincide with the corresponding equation in \cite{Carbotte1}:
\beg\label{RealDLT}
\begin{split}
&Z(\omega)\Delta(\omega)=\pi T\sum\limits_{m=-\infty}^\infty\frac{D_{nm}\Delta_m}{\sqrt{\omega_m^2+\Delta_m^2}}\\&+i\pi A\left\{\frac{\Delta(\omega-\Omega)\left[n_B(\Omega)+n_F(\Omega-\omega)\right]}{[(\omega-\Omega+i0)^2-\Delta^2(\omega-\Omega)]^{1/2}}\right.\\&\left.
+\frac{\Delta(\omega+\Omega)\left[n_B(\Omega)+n_F(\Omega+\omega)\right]}{[(\omega+\Omega+i0)^2-\Delta^2(\omega+\Omega)]^{1/2}}\right\}.
\end{split}
\en
Here parameter $A=\lambda\Omega/2$, $n_B(\omega)=[\exp(\omega/T)-1]^{-1}$ is the Bose distribution function and $n_F(\omega)=[\exp(\omega/T)+1]^{-1}$ is the Fermi distribution function. 
Conversely, given the presence of the self-energy correction due to Kondo lattice effects, the equation for the function $Z(\omega)$ becomes
\beg\label{RealZw}
\begin{split}
&Z(\omega)=1+\frac{a_{\textrm{sb}}^2}{(\omega+i0)^2-\veps_f^2}\\&+\frac{i\pi T}{\omega+i0}\sum\limits_{m=-\infty}^\infty\frac{D_{nm}\omega_m}{\sqrt{\omega_m^2+\Delta_m^2}}\\&+i\pi A\left\{\frac{(\omega-\Omega)\left[n_B(\Omega)+n_F(\Omega-\omega)\right]}{[(\omega-\Omega+i0)^2-\Delta^2(\omega-\Omega)]^{1/2}}\right.\\&\left.
+\frac{(\omega+\Omega)\left[n_B(\Omega)+n_F(\Omega+\omega)\right]}{[(\omega+\Omega+i0)^2-\Delta^2(\omega+\Omega)]^{1/2}}\right\}, \\
\end{split}
\en
Expressions (\ref{RealDLT},\ref{RealZw}) give correct analytic continuation of the equations (\ref{EliKLFin}) in the upper half plane, so that $Z(w)$ is analytic in the upper half-plane and has multiple poles given by $\pm n\Omega+\Delta(\omega\pm n\Omega)$ and $\veps_f$ in the lower half plane of complex $\omega$.

\begin{figure}
\includegraphics[width=0.8\linewidth]{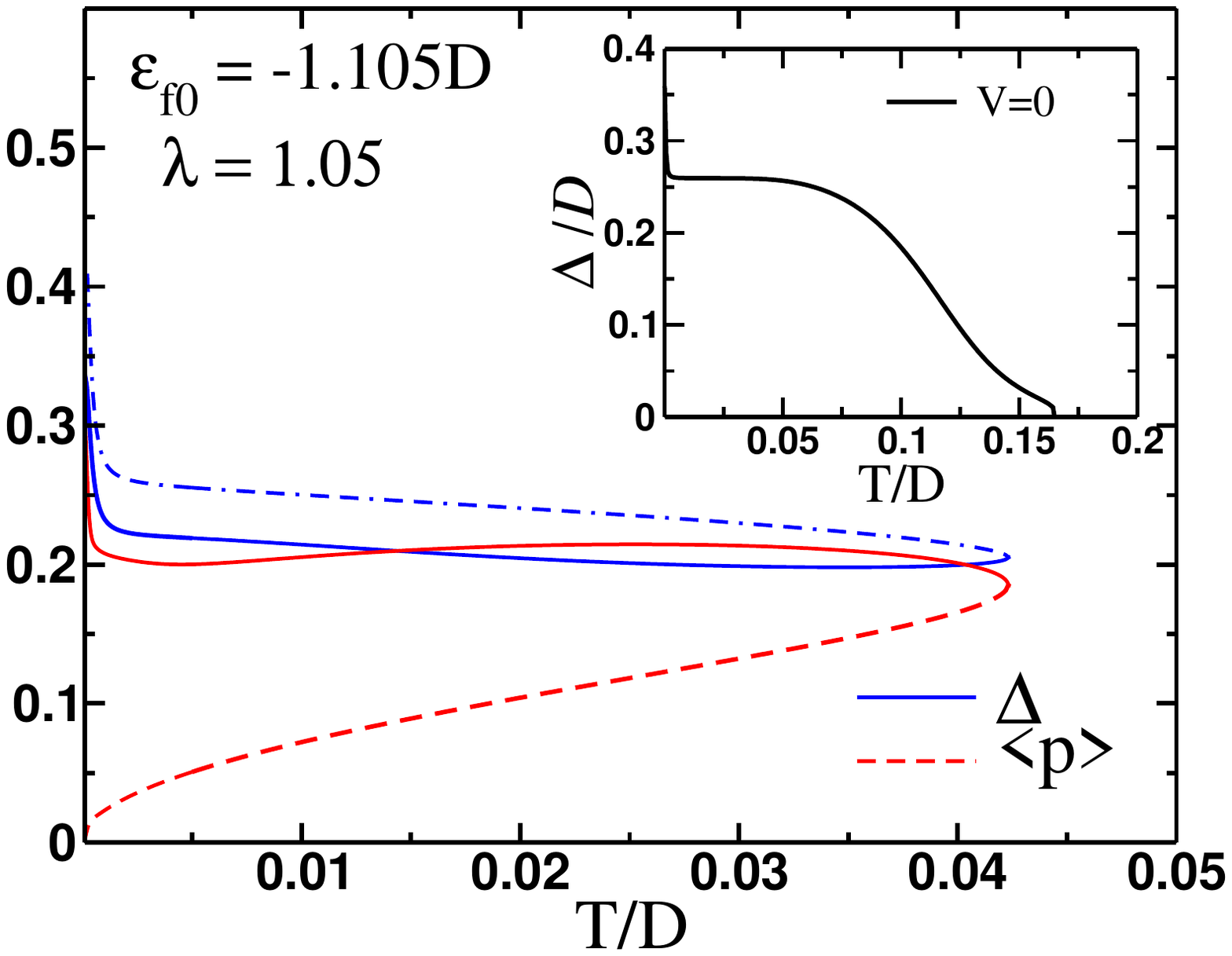}
\caption{Plot of the temperature dependence of the pairing gap $\Delta$ (units of ${\cal D}$) and slave-boson amplitude $\langle p\rangle$ when $\lambda=1.05$. The position of the bare $f$-level has been chosen in the region when two solutions exist. The solid lines correspond to the solution with lower free energy. Inset: temperature dependence of $\Delta$ in the absence of hybridization.}
\label{Fig-DLTvsT}
\end{figure}

In Fig. \ref{Fig-DLTvsT} we show the results of the iterative solution of the Eliashberg equations (\ref{RealDLT},\ref{RealZw}) for $\lambda=1.05$ and $\eps_{f0}=-1.105D$. As we have already discussed above, as temperature is lowered out solutions indicate that there is a 'first-order-like' transition into a heavy-fermion state, when $\langle p \rangle$ acquires a finite value, while the pairing gap decreases slightly from its value for $V=0$ (see inset in Fig. \ref{Fig-DLTvsT}). Note that as temperature is lowered the system will go into an unscreened state at $T^*\approx 0.015{\cal D}$ and then into a coherent state again as the temperature approaches zero. As far as we know, this is the first example when the development of the many-body coherence exhibits such a 're-entrant' behavior. 

\section{Conclusions}
In this paper we have studied the onset of the many-body coherence in the Kondo lattice in the presence of the electron-phonon interaction. We have derived a system of nonlinear equations using the saddle-point approximation. When the electron-phonon coupling is strong enough, we find two solutions which have different values of the free energy. This result implies that at strong coupling the the many body coherence emerges abruptly and upon further decrease in temperature becomes re-entrant. Our work provides the first example when the Kondo screening in the Kondo lattice exhibits a re-entrant behavior similar to the superconducting transition in superconducting alloys contaminated with Kondo impurities. Our results can also be used in the context of the related problem: an interplay of the Kondo effect and strong coupling superconductivity in diluted magnetic alloys. We leave this problem for the future studies.

\section{Acknowledgments}
We would like to thank Emil Yuzbashyan and Ilya Vekhter for very useful discussions.
This work was financially supported by the National Science Foundation grant NSF-DMR-2002795 (S.A. and M.D.) This project was started during the Aspen Center of Physics 2023 Summer Program on '\emph{New Directions on Strange Metals in Correlated Systems}', which was supported by the National Science Foundation Grant No. PHY-2210452.

\begin{appendix}
\section{Effective bosonic action}\label{EffectiveAction}
At the level of the saddle-point apprxomation, the corresponding saddle-point equations can now be obtained by find an extremum of an action
\beg\label{SeffMatsubara}
\begin{split}
S_{\textrm{eff}}&=\nu_FT\sum\limits_{nl}\left[\Phi_{n+l}^*D_{l}^{-1}\Phi_n+\Sigma_{n+l}D_{l}^{-1}\Sigma_n\right]\\&+\frac{2}{T}\left(\langle{p}\rangle^2-\frac{1}{2}\right)(\veps_f-\eps_{f0})-\textrm{Tr}\log\hat{\cal M}.
\end{split}
\en
Here $\nu_F$ is the single particle density of states at the Fermi level, $D_l^{-1}$ is the Matsubara frequency component of $D^{-1}(\tau-\tau')$ and
\begin{widetext}
\beg\label{Seff}
\begin{split}
&\hat{\cal M}(\bk,\omega_n)=\left[
\begin{matrix}
-i\left(\omega_n+\Sigma_n\right)+\xi_{\bk} & \Phi_n & V\langle{p}\rangle & 0 \\
\Phi_n & -i\left(\omega_n+\Sigma_n\right)-\xi_{\bk} & 0 & -V\langle{p}\rangle \\
V\langle{p}\rangle & 0 & -i\omega_n+\veps_f & 0 \\
0 & -V\langle{p}\rangle & 0 &  -i\omega_n-\veps_f
\end{matrix}
\right], 
\end{split}
\en
where $\xi_{\bk}=\eps_\bk-\mu$. 
\end{widetext}
Note that we are not including the bosonic fields $\chi_n$ \cite{EmilEli1} in our considerations since they are only nonzero in the absence of the particle-hole symmetry. These fields will be neglected in our discussion in the main text. Then we can choose bosonic fields $\Phi_n$ as purely real. 
In view of this approximation, for the last term in (\ref{Seff}) it obtains
\beg\label{LogM}
\begin{split}
&\textrm{Tr}\log\hat{\cal M}=\sum\limits_{n\bk}\log\left\{\left(\omega_n^2+\veps_f^2\right)\right.\\&\left.\times[(\omega_n(1+\sigma_n)+\Sigma_n)^2+\Phi_n^2+(\xi_\bk-\veps_f\sigma_n)^2]\right\}\\&\approx2\pi\nu_F\sum\limits_{n}\sqrt{[\omega_n(1+\sigma_n)+\Sigma_n]^2+\Phi_n^2}
\end{split}
\en
and we introduced $\sigma_n=a_{\textrm{sb}}^2/(\omega_n^2+\veps_f^2)$, $a_{\textrm{sb}}=V\langle p\rangle$. Although integration over $\xi_\bk$ is approximate, it allows us to use the pseudospin representation for the Migdal-Eliashberg theory to see how the presence of the underlying Kondo lattice affects the superconducting properties. Indeed, introducing variables
\beg\label{SnxSnz}
\begin{split}
S_n^x&=\frac{\Phi_n}{[\omega_n(1+\sigma_n)+\Sigma_n]^2+\Phi_n^2}, \\
S_n^z&=\frac{\omega_n(1+\sigma_n)+\Sigma_n}{[\omega_n(1+\sigma_n)+\Sigma_n]^2+\Phi_n^2}.
\end{split}
\en
Then the free energy at the saddle point (excluding the Kondo lattice part) we readily obtain
the expression (\ref{KLSpins}) in the main text. 

Equation which determines the position of the chemical potential can be conveniently written in terms of the single-particle propagators. The corresponding expressions for them are given by:
\begin{widetext}
\beg\label{GccGff}
\begin{split}
{\cal G}_{\textrm{cc}}(i\omega_m,\bk)&=\frac{a_{\textrm{sb}}^2-\left(i\omega_m+\veps_f\right)(i\omega_m+
i\Sigma_m+\xi_\bk)}{\left(i\omega_m+\veps_f\right)\left\{
[\omega_m^2Z_m^2+(\xi_\bk-\veps_f\sigma_m)^2+|\Phi_m|^2]\right\}}, \\
{\cal G}_{\textrm{ff}}(i\omega_m,\bk)&=\frac{-(i\omega_m+\veps_f)[(\omega_m+
\Sigma_m)^2+\xi_\bk^2+|\Phi_m|^2]-a_{\textrm{sb}}^2(i\omega_m+i\Sigma_m-\xi_\bk)}{\left(\omega_m^2+\veps_f^2\right)\left\{
\omega_m^2Z_m^2+(\xi_\bk-\veps_f\sigma_m)^2+|\Phi_m|^2]\right\}}.
\end{split}
\en
In the limit when $\Phi_n=\Sigma_n=0$, we readily recover the familiar expressions
\beg\label{FamExpr}
\begin{split}
&{\cal G}_{\textrm{cc}}^{(0)}(i\omega_m,\bk)=\frac{i\omega_m-\veps_f}{(i\omega_m-\veps_f)(i\omega_m-\xi_\bk)-a_{\textrm{sb}}^2},\\ 
&{\cal G}_{\textrm{ff}}^{(0)}(i\omega_m,\bk)=\frac{i\omega_m-\xi_\bk}{(i\omega_m-\veps_f)(i\omega_m-\xi_\bk)-a_{\textrm{sb}}^2}.
\end{split}
\en
Lastly, total particle number (per spin) is
\beg\label{avn}
\begin{split}
&T\sum\limits_{\omega_m}\sum\limits_{\bk}\left[{\cal G}_{\textrm{cc}}(i\omega_m,\bk)+{\cal G}_{\textrm{ff}}(i\omega_m,\bk)\right]e^{i\omega_m0+}=n_{\textrm{c}}+n_{\textrm{f}}={n}_{\textrm{tot}}.
\end{split}
\en
Since we consider ${n}_{\textrm{tot}}=0.875$ to be fixed, while $a_{\textrm{sb}}$ and $\veps_f$ as the free parameters of the theory, it will be more convenient to re-write the second equation, so that the convergence of the integral over $\bk$ is facilitated. 
Next we introduce functions
\beg\label{calFs}
\begin{split}
&{\cal F}_1[{\vec \Sigma},{\vec \Phi}]=T\sum\limits_{\omega_m}\sum\limits_{\bk}\frac{V^2[\omega_m(\omega_m+\Sigma_m)-\veps_f\xi_\bk+a_{\textrm{sb}}^2]}{\left(\omega_m^2+\veps_f^2\right)\left\{
(\omega_m^2+\Delta_m^2)Z_m^2+(\xi_\bk-\veps_f\sigma_m)^2\right\}}, \\
&{\cal F}_2[{\vec \Sigma},{\vec \Phi}]=T\sum\limits_{\omega_m}\sum\limits_{\bk}
\frac{\veps_f[(\omega_m+\Sigma_m)^2+\xi_\bk^2+Z_m^2\Delta_m^2]-\xi_\bk a_{\textrm{sb}}^2}{\left(\omega_m^2+\veps_f^2\right)\left\{
(\omega_m^2+\Delta_m^2)Z_m^2+(\xi_\bk-\veps_f\sigma_m)^2\right\}}, \\
&{\cal F}_3[{\vec \Sigma},{\vec \Phi}]=T\sum\limits_{\omega_n\bk}\left\{{\cal G}_{\textrm{cc}}(i\omega_m,\bk)+{\cal G}_{\textrm{ff}}(i\omega_m,\bk)\right\}e^{i\omega_n0+}.
\end{split}
\en
to facilitate the convergence of the momentum summations. 
\end{widetext}

\end{appendix} 
 

\end{document}